# Pairs and groups of asteroids with nearest orbits in known families

Rosaev A.E.

*NPC Nedra, Yaroslavl, Russia*

Abstract: The problem of origin and age of asteroid families studied very intensively. First of all youngest families are interesting due to possibility of the reconstruction collisional history. But in oldest families present objects with very close orbits some of them are products of recent collisions. The search for and studying of dynamics of such pairs of orbits is a main aim of the present paper. In result, two new young groups of minor planets are detected as well as some new members of known families.

The problem of origin and age of asteroid families studied in many publications. First of all youngest and compact minor planets clusters are interesting due to possibility of the reconstruction collisional history. The search for of such pairs or groups of orbits is a main aim of the present paper. The list of orbital elements of 415000 asteroids with permanent numbers from the catalogue [1] on data 15 Dec 2014 is used. The method of selection of nearest orbits is similar to used in **[2]**.

Due to very high sensitivity to variation of angular elements, this method has an advantage in detecting first of all **low velocity and most recent** breakups. We use osculating orbital elements instead proper elements by similar reason – our target is search for **low velocity and most recent** breakups

The clusters of close orbits with 3 or more members have a particular interest**.** In this case, we can say about new compact young family, was born in a low-velocity breakup. First of all, point of attention extended group of orbits close to minor planet 61510 (2000 QF55) **(table 1)**. It is moving near 3:1 mean motion resonance with Jupiter.

**Table 1.** Orbital elements of **61510** (2000 QF55) family

| Object | | ω | Ω | *i* | *e* | *a* |
|---|---|---|---|---|---|---|
| 61510 | 2000 QF55 | 330.123898 | 33.328532 | 3.782726 | 0.19451320 | 2.561987 |
| 244489 | 2002 TD39 | 330.100764 | 33.509563 | 3.483690 | 0.19081589 | 2.533620 |
| 64877 | 2001 YH64 | 329.887255 | 34.234614 | 3.962953 | 0.18974534 | 2.516288 |

One interesting group was found in Eos family (table 2).

**Table 2**. Orbital elements of **72451 (**2001 DW5) group (Eos family)

| Object | | ω | Ω | *i* | *e* | *a* |
|---|---|---|---|---|---|---|
| 274613 | 2008 TJ59 | 242.018157 | 30.415983 | 10.440521 | 0.06441258 | 3.022184 |
| 72451 | 2001 DW5 | 242.116590 | 30.664965 | 10.342423 | 0.06348877 | 2.996550 |
| | 2008 UM306 | 242.813131 | 30.403619 | 10.589472 | 0.06535584 | 2.985170 |

A special search for new members of very young families (Emilkowalski, Lucascavin, 1992YC2) detected in **[2]** gives one new member (256124 2006 UK337) of Emilkowalski cluster (table 3) and one new member (2009 VZ4) of 1992YC2 cluster (table 4). It gives a hope to improve estimation of age of this minor planet association.

Table 3. Orbital elements of Emilkowalski cluster with new member 256124 (2006 UK337)

| Object | | Ω | ω | i | e | a |
|---|---|---|---|---|---|---|
| 14627 | Emilkowalski | 41.557876 | 44.282158 | 17.733873 | 0.15046022 | 2.598322 |
| 126761 | 2002 DW10 | 41.329414 | 44.660678 | 17.748749 | 0.15101571 | 2.587832 |
| 224559 | 2005 WU178 | 42.238339 | 44.951268 | 17.751624 | 0.15143797 | 2.587319 |
| 256124 | 2006 UK337 | 42.415539 | 43.229893 | 17.745496 | 0.15023856 | 2.598752 |

Table 4. Orbital elements of 1992YC2 cluster with new member (2009 VZ4)

| Objects | | ω | Ω | *i* | *e* | *a* |
|---|---|---|---|---|---|---|
| 16598 | (1992YC2) | 286.889923 | 105.578820 | 1.627700 | 0.21921664 | 2.620864 |
| 218697 | (2005TT99) | 286.531207 | 105.908155 | 1.626207 | 0.21805414 | 2.621559 |
| 190603 | (2000UV80) | 286.491882 | 106.251948 | 1.629244 | 0.21968573 | 2.619551 |
|  | (2009 VZ4) | 286.663815 | 104.941615 | 1.312550 | 0.22096234 | 2.619935 |

Based on present youngest families mean age estimation (1992YC2 **[2]**, Emilkowalski **[2]**, Datura **[4]**, Rheinland **[5]**, Veritas **[3]**) we note that main difference in pericentre longitude depends on (mean) age of each family (Fig.1).

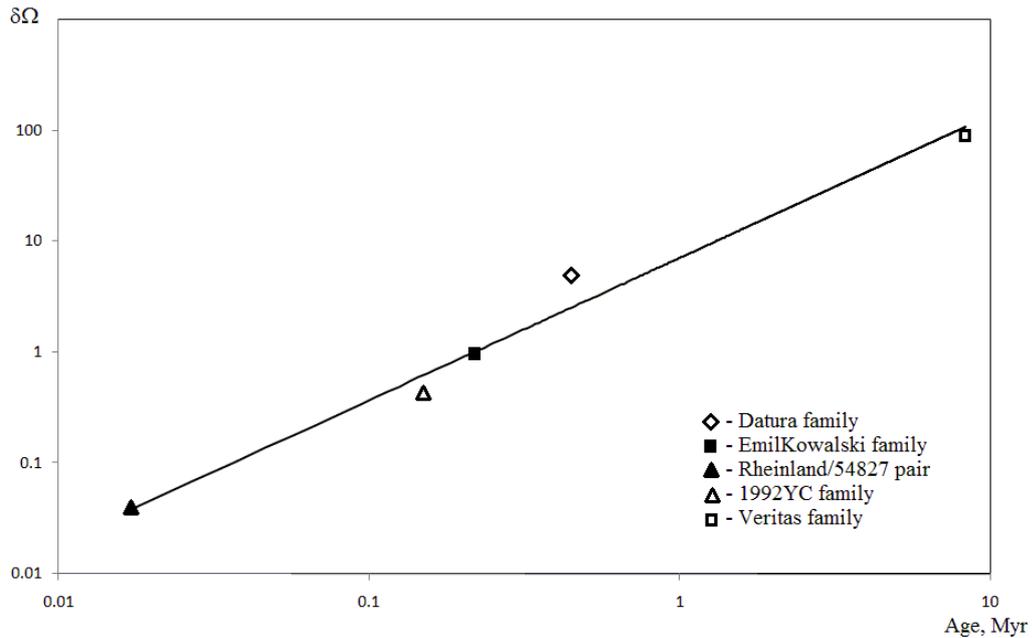

Fig. 1. δΩ on age dependence for young asteroid families

This dependence is possible to explain in according with model minimal velocity of breakup [2]. First of all, at the epoch of the origin of asteroids as an independent objects, relative velocity was minimal, after that it increase due to differential planetary perturbations. Secondly, at the epoch of breakup difference between node longitudes of objects was close to zero.

Two pairs of very close orbits were found in Karin cluster **(table 5)**. Note that, angular elements in these pairs are significantly differ one from another, it means different epochs of origin.

Table 5. Pairs of close orbits in Karin family

| N | Object | ω | Ω | *i* | *e* | *a* |
|---|---|---|---|---|---|---|
| 1 | 304338 | 228.842584 | 183.894988 | 1.794079 | 0.06006250 | 2.86496300 |
|  | 27663 | 228.620518 | 183.440275 | 1.812611 | 0.06082590 | 2.86713300 |
| 2 | 346031 | 202.650047 | 215.271946 | 1.283998 | 0.05926005 | 2.86614400 |
|  | 143155 | 202.198601 | 215.310605 | 1.282446 | 0.05779043 | 2.86672600 |

Few words is necessary to say about founded pairs of close orbits. The pairs of nearest orbits are non-homogenous distributed in families. There are very small numbers of pairs in Maria and Eunomia and Higea families. In contrast, many pairs were found in Flora family **(table 6)**, Nisa and Massalia family. However, as well as we use osculating elements, some interlopers may be present in table 5.

Table 6. Pairs of close orbits in Flora family

| N | Objects | ω | Ω | i | e | a |
|---|---------|-----|-----|-----|-----|-----|
| 1 | 38707 | 176.982580 | 207.229277 | 5.928168 | 0.11564556 | 2.277854 |
|   | 32957 | 176.968946 | 206.997861 | 5.928747 | 0.11596370 | 2.277985 |
| 2 | 44645 | 190.072919 | 253.446524 | 5.731073 | 0.07832158 | 2.320515 |
|   | 10484 | 190.051270 | 253.315340 | 5.729344 | 0.07841743 | 2.320624 |
| 3 | 153536 | 33.883724 | 184.917552 | 6.835092 | 0.06852059 | 2.331234 |
|   | 43649 | 34.133581 | 184.816272 | 7.194358 | 0.06633799 | 2.337967 |
| 4 | 169507 | 239.339927 | 169.120942 | 4.882676 | 0.13458052 | 2.284579 |
|   | 14759 | 239.743496 | 169.280259 | 4.879014 | 0.13418490 | 2.284498 |
| 5 | 172094 | 229.745213 | 348.418131 | 4.672833 | 0.08045507 | 2.262899 |
|   | 164227 | 229.686161 | 348.473969 | 4.599663 | 0.08226056 | 2.269051 |
| 6 | 180831 | 291.941388 | 5.924105 | 6.572076 | 0.13113504 | 2.302954 |
|   | 43760 | 291.808257 | 5.845339 | 6.763166 | 0.13326735 | 2.294414 |
| 7 | 182259 | 100.868849 | 48.342152 | 5.837627 | 0.10073458 | 2.247000 |
|   | 2897 | 101.138131 | 48.331919 | 5.837617 | 0.09999886 | 2.247599 |
| 8 | 193865 | 145.271544 | 201.747150 | 6.533669 | 0.14047515 | 2.308341 |
|   | 6085 | 145.493582 | 201.563056 | 6.645161 | 0.13873260 | 2.306821 |
| 9 | 194083 | 145.547708 | 294.422585 | 2.230352 | 0.07403302 | 2.340753 |
|   | 92652 | 145.465717 | 294.402089 | 2.231697 | 0.07390463 | 2.341015 |
| 10 | 209570 | 4.608485 | 70.069065 | 5.983802 | 0.11183813 | 2.280948 |
|   | 21509 | 4.224905 | 70.179929 | 5.981963 | 0.11235157 | 2.281033 |
| 11 | 211118 | 203.584184 | 29.869763 | 6.055903 | 0.12127499 | 2.291480 |
|   | 74913 | 203.544004 | 29.509133 | 6.056084 | 0.12066407 | 2.288784 |
| 12 | 217266 | 197.064358 | 87.160814 | 3.840546 | 0.11504732 | 2.236775 |
|   | 180906 | 197.232323 | 87.163312 | 3.840663 | 0.11517782 | 2.236268 |
| 13 | 220103 | 350.743982 | 21.622843 | 7.129577 | 0.12567246 | 2.297522 |
|   | 28277 | 350.728541 | 21.780279 | 6.989241 | 0.12776201 | 2.307853 |
| 14 | 220143 | 70.456501 | 261.065579 | 5.514047 | 0.13165174 | 2.324131 |
|   | 54041 | 70.335720 | 261.095200 | 5.515646 | 0.13271353 | 2.323197 |
| 15 | 228747 | 172.127724 | 273.099249 | 3.690776 | 0.09430342 | 2.250262 |
|   | 11842 | 172.346045 | 272.844049 | 3.688654 | 0.09440498 | 2.249927 |
| 16 | 229056 | 252.108008 | 12.387186 | 3.289355 | 0.10157665 | 2.279529 |
|   | 17198 | 252.473996 | 12.155613 | 3.285774 | 0.10171839 | 2.279591 |
| 17 | 253736 | 74.917257 | 254.289545 | 5.863425 | 0.14758679 | 2.311927 |
|   | 11575 | 75.201568 | 254.142320 | 5.701142 | 0.14924919 | 2.308001 |
| 18 | 276353 | 36.076417 | 23.420605 | 4.252699 | 0.15660043 | 2.337329 |
|   | 57202 | 36.095501 | 23.432219 | 4.250974 | 0.15623187 | 2.337511 |
| 19 | 282206 | 124.608308 | 268.517753 | 6.693586 | 0.11129367 | 2.308879 |
|   | 165389 | 124.741579 | 268.514642 | 6.696536 | 0.11091755 | 2.308498 |
| 20 | 295745 | 181.275022 | 182.188328 | 4.390261 | 0.16296354 | 2.176668 |
|   | 44620 | 180.916278 | 182.017570 | 4.398675 | 0.16420201 | 2.176320 |
| 21 | 297862 | 10.577786 | 101.675490 | 3.024848 | 0.16265898 | 2.330985 |
|   | 244175 | 10.616932 | 101.891561 | 3.309244 | 0.16581033 | 2.346168 |
| 22 | 334916 | 178.177263 | 320.373626 | 3.737290 | 0.08502729 | 2.186524 |
|   | 21436 | 178.175708 | 320.376576 | 3.737201 | 0.08506250 | 2.186466 |
| 23 | 392940 | 1.016127 | 359.711935 | 2.813416 | 0.18457083 | 2.301057 |
|   | 303310 | 0.930880 | 359.464386 | 2.780224 | 0.18165372 | 2.296256 |
| 24 | 404118 | 114.233701 | 129.870461 | 2.796669 | 0.14320857 | 2.217332 |
|   | 355258 | 114.245194 | 129.870604 | 2.796669 | 0.14315843 | 2.217367 |
| 25 | 412065 | 137.357261 | 141.019013 | 7.230954 | 0.13119306 | 2.216828 |
|   | 11677 | 137.402066 | 141.299782 | 7.229723 | 0.12996014 | 2.217464 |

Particular interest have pairs, related with recently detected [6] comet Gibbs family. It is important due to observed activity P/2012 F5 (Gibbs) can be explained by recent small body impact [7]. We find one new member (**389622** 2011HU90) of Gibbs association (table 7). On the other hand, minor planet 140429 has elements, significantly far from other members of the group and its belonging to cluster remains open. The mean difference in orbital elements of Gibbs association is relatively large than in other families. One possible reason is that relative velocity of breakup in this case is large, in spite this family is young (1.5 Myr [6]).

Table 7. Asteroids belonging to the Gibbs cluster.

| Objects | | ω | Ω | *i* | *e* | *a* |
|---|---|---|---|---|---|---|
| 20674 | 1999 VT1 | 198.182456 | 204.515556 | 9.949356 | 0.03637542 | 3.004393 |
| 140429 | 2001 TQ96 | 161.955158 | 227.105282 | 9.598366 | 0.05608882 | 3.001735 |
| 177075 | 2003 FR36 | 212.658396 | 184.872593 | 10.321428 | 0.02434394 | 3.005203 |
| 249738 | 2000 SB159 | 210.956367 | 189.880506 | 10.224040 | 0.02819309 | 3.004990 |
| 257134 | 2008 GY132 | 213.372956 | 182.716426 | 10.372597 | 0.02052292 | 3.005724 |
| 321490 | 2009 SH54 | 217.930008 | 182.550232 | 10.374016 | 0.01893629 | 3.005388 |
| 341222 | (2007RT138) | 230.463027 | 187.427610 | 10.267488 | 0.02872199 | 3.007247 |
| **389622** | (2011HU90) | 229.595009 | 187.061339 | 10.268341 | 0.02362280 | 3.004487 |
| | (2002TF325) | 230.190581 | 183.142128 | 10.345954 | 0.01915150 | 3.005350 |
| | P/2012 F5 | 274.329067 | 216.987096 | 9.7282099 | 0.04404619 | 3.007766 |

The dependence inclination on ascending node longitude for members of Gibbs cluster is very close to linear (Fig.2). It is interesting, that similar dependence take place for closest pairs of orbits in Coronis family (fig.3). In contrary, closest pairs in Flora family (table 6) distributed randomly in (*i*, Ω) coordinates.

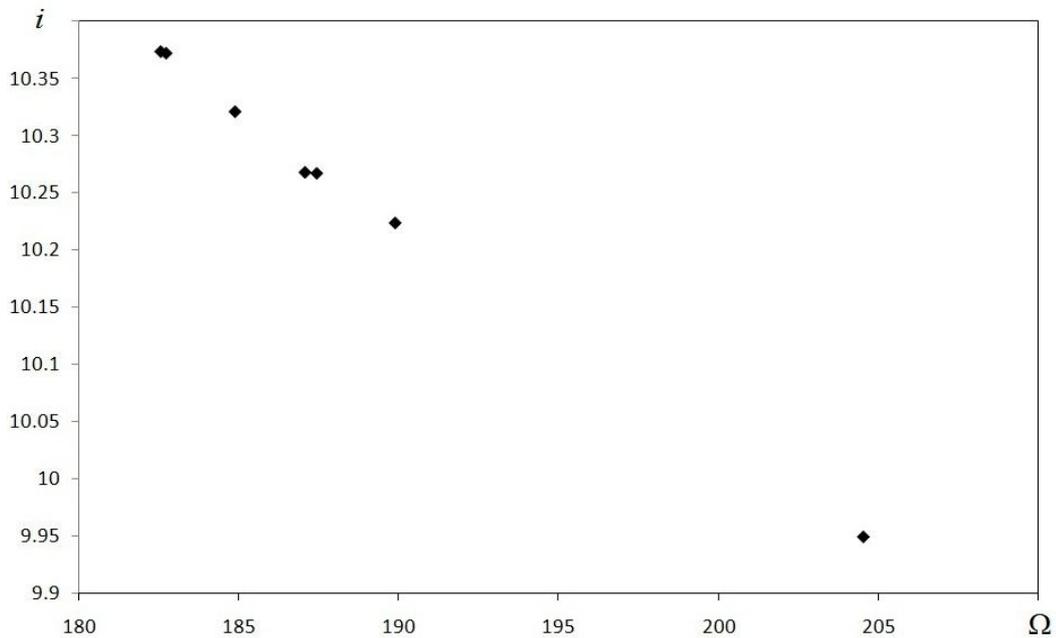

Fig.2. The positions members of Gibbs family in (*i*, Ω) coordinates

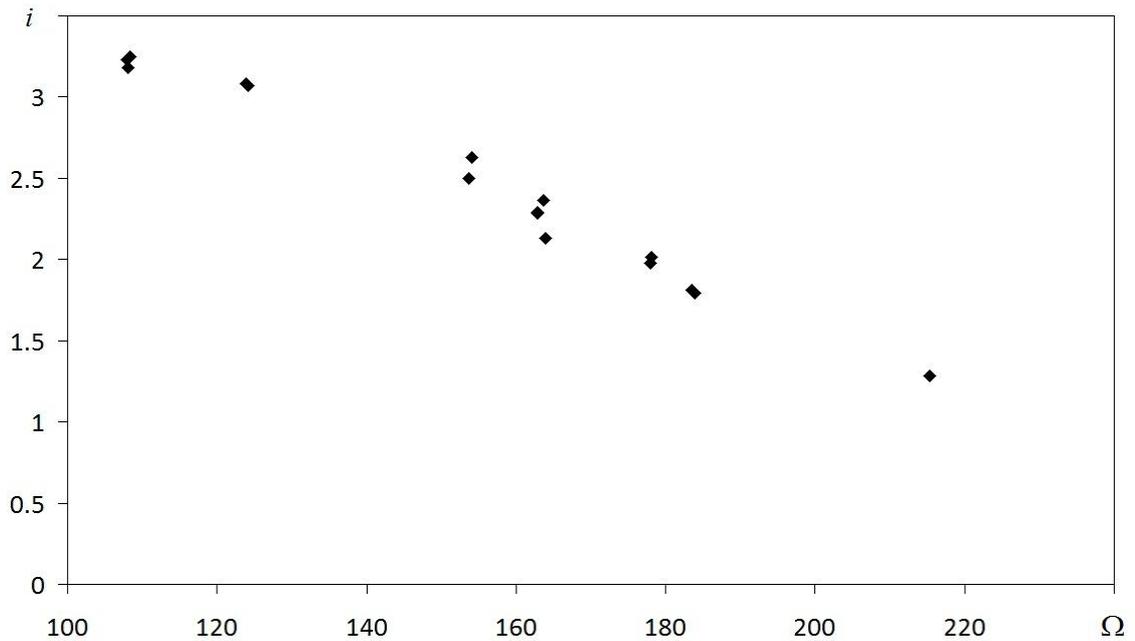

Fig.3. The positions of most close pairs of orbits in Coronis family in ($i$, $\Omega$) coordinates

Finally we can conclude that process of fragmentation of minor planets in families is continued. In some cases of most recent events, the reconstruction of origin of asteroids on close orbits is possible in model of low-velocity breakup. Origin of some families was not one-momentum. Other result of this paper is a detection a new members of young asteroid groups. New compact groups related with asteroids **61510** (2000 QF55) and **72451 (**2001 DW5) required future studying.